\documentclass[usenatbib]{mn2e}

\usepackage{amsmath,amssymb, graphics}
\usepackage{pgf,tikz}
\usepackage{bm}
\usepackage{graphicx}
\usepackage{epstopdf}
\usepackage{float}
\usepackage{mathrsfs}

\newcommand{\ag}{\alpha}

\newcommand{\dg}{\delta}

\newcommand{\Om}{\Omega}

\newcommand{\der}{\text{d}}

\newcommand{\cs}{c_{\rm s}}

\newcommand{\eps}{\epsilon}

\newcommand{\del}{\nabla}
\newcommand{\rg}{\rho}

\newcommand{\bT}{{\bf T}}
\newcommand{\bdel}{{\bm \nabla}}
\newcommand{\Del}{\Delta}

\newcommand{\vphi}{\varphi}
\newcommand{\vr}{{\bm {\hat r}}}

\newcommand{\xir}{\xi_r}
\newcommand{\xip}{\xi_\perp}

\newcommand{\rgc}{\rho_{\rm c}}
\newcommand{\rgo}{\rho_{\rm o}}
\newcommand{\rga}{\rho_{\rm a}}
\newcommand{\Rc}{R_{\rm c}}
\newcommand{\gc}{g_{\rm c}}

\newcommand{\epsmax}{\epsilon_{\rm max}}

\newcommand{\bu}{{\bf u}}
\newcommand{\baru}{\bar u}
\newcommand{\barupeak}{{\bar u}_{\rm peak}}
\newcommand{\upeak}{u_{\rm peak}}
\newcommand{\barusurf}{{\bar u}_{\rm surf}}

\newcommand{\ucrit}{u_{\rm crit}}

\newcommand{\bcdot}{{\bm \cdot}}

\newcommand{\be}{\begin{equation}}
\newcommand{\ee}{\end{equation}}
\newcommand{\ba}{\begin{align}}
\newcommand{\ea}{\end{align}}

\begin{document}


\title[Triaxial Deformation of Rocky Exoplanets] 
{Triaxial Deformation and Asynchronous Rotation of Rocky Planets in
the Habitable Zone of Low-Mass Stars}
%\title[Maximal Triaxiallity of Rocky Exoplanets]
%{Geophysical Constraints on the Maximal Triaxiallity of Rocky Exoplanets}

\author[J. J. Zanazzi and Dong Lai]{J. J. Zanazzi$^{1}$\thanks{Email: jjz54@cornell.edu}, and Dong Lai$^{1}$ \\
$^{1}$Cornell Center for Astrophysics and Planetary Science, Department of Astronomy, Cornell University, Ithaca, NY 14853, USA}

\maketitle
%DL: rewrite
\begin{abstract}
Rocky planets orbiting M-dwarf stars in the habitable zone tend to be
driven to synchronous rotation by tidal dissipation, potentially
causing difficulties for maintaining a habitable climate on the
planet.  However, the planet may be captured into asynchronous
spin-orbit resonances, and this capture may be more likely if the planet has a  sufficiently large intrinsic triaxial
deformation. We derive the analytic expression for the maximum
triaxiality of a rocky planet, with and without a liquid envelope, as a
function of the planet's radius, density, rigidity and critical strain
of fracture. The derived maximum triaxiality is consistent with the
observed triaxialities for terrestrial planets in the solar system, and
indicates that rocky planets in the habitable zone of M-dwarfs can in
principle be in a state of asynchronous spin-orbit resonances.
\end{abstract}


\section{Introduction}
\label{sec:Intro}

With current technology, we may detect rocky exoplanets in the
habitable zone (HZ) of M-dwarf
stars \citep{CharbonneauDeming(2007),Shields(2016)}.  Indeed, the
TRansiting Planets and PlanestIsimals Small Telescope (TRAPPIST)
survey has already discovered several potentially habitable planets
around the low-mass ($0.08 M_\odot$) star TRAPPIST-1 \citep{Gillon(2016),Gillon(2017)}, and radial velocity
measurements have revealed the earth-massed planet Proxima Centauri b
in the HZ around the closest star to our sun
\citep{Anglada-Escude(2016)}.
%Such discoveries will increase dramatically with the release of the
%Transiting Exoplanet Survey Satellite (TESS, \citealt{Ricker(2014)}), as 
%statistics on the planetary architectures which Kepler mission unveiled argue 
Statistics of planets discovered by the Kepler mission suggests that 
$\sim 50 \%$ of stars with effective temperatures
cooler than $4000^\circ \, \text{K}$ have earth-sized 
planets \citep{DressingCharbonneau(2013),DressingCharbonneau(2015),MortonSwift(2014)},
and $\sim 20 \%$ of these cool stars have rocky planets
in the HZ \citep{MortonSwift(2014),DressingCharbonneau(2015)}.
 
Because the HZ of M-dwarfs is located at a small orbital semi-major axis
($a \lesssim 0.1$~AU), planets in this region are often expected to be
in a state of tidally synchronized rotation. This could potentially
create difficulties for maintaining a habitable climate over the
lifetime of the planet, and even lead to atmosphere collapse
(e.g., \citealt{Kasting(1993),Joshi(1997),Kite(2011),HengKopparla(2012),Yang(2013),Kopparapu(2016),Turbet(2016)}).
%DL: Add REFs below:
%Kasting et al: http://adsabs.harvard.edu/abs/1993Icar..101..108K
%Joshi et al:   http://adsabs.harvard.edu/abs/1997Icar..129..450J
%Kite et al:    http://adsabs.harvard.edu/abs/2011ApJ...743...41K
%Heng et al:    http://adsabs.harvard.edu/abs/2012ApJ...754...60H
Thermal tide associated with a sufficiently massive atmosphere
can in principle drive the planet's rotation away from synchronicity.
This is the case for Venus \citep{GoldSoter(1969),IngersollDobrovolskis(1978)},
and may also operate for planets in the HZ around stars more massive than
$0.5M_\odot$ \citep{Leconte(2015)}.
%DL: add REFs below
% Leconte et al.~2015, Science, 347, 632
% (The other REFs can be found in the Leconte et al paper)
Another possibility to avoid tidal locking is the planet
retains a small orbital eccentricity, while spin is captured into 
a non-synchronous resonance (such as $3:2$) with the orbit during 
spindown, as in the case of Mercury \citep{GoldreichPeale(1966),GoldreichPeale(1968)}.

%Planets in the HZ of M-dwarfs will have strong tidal interactions with
%the host star, which will effect the plant's rotation
%\citep{GoldreichSoter(1966)}.  The most probable end state for the
%planet's rotation is the planetary spin becoming synchronized, or
%capture into a 3:2 spin-orbit resonance
%\citep{GoldreichPeale(1966),GoldreichPeale(1968),MurrayDemott(2000)}.
%These two spin states have dramatically different climates for planets
%in the HZ of M-dwarf stars
%\citep{Yang(2013),Kopparapu(2016),Ribas(2016),Turbet(2016)}.

A critical parameter for determining if a planet is susceptible to be
captured into a spin-orbit resonance is its intrinsic triaxiality [see
  Eq.~\eqref{eq:eps}].  This triaxiality is sustained by the rigidity
the rocky planet, and determines the strength of the torque
keeping the planet in resonance. For the simplest frequency-independent rheologies, this resonant triaxial torque must overcome the
dissipative tidal torque working to drive the planet toward
synchronization \citep{GoldreichPeale(1966),GoldreichPeale(1968),MurrayDemott(2000)}.
 With frequency dependent rheologies \citep{Makarov(2012),Efroimsky(2012)}, spin-orbit resonant capture may occur without the resonant triaxial torque due to the behavior of the tidal torque near spin-orbit resonances \citep{Ribas(2016),Bartuccelli(2017)}.  However, the resonant triaxial torque is often necessary for spin-orbit resonant capture, even with frequency-dependent rheologies (see Fig. 4 of \citealt{Ribas(2016)}).
The main goal of this paper is to calculate the maximum triaxial deformation
a rocky planet (with and without a fluid envelope) can sustain, as a function
of its physical and material properties (density, size, elastic rigidity and the 
critical strain for fracture), and to evaluate the possibility of 
resonant spin-orbit capture of planets in the HZ of M-dwarfs.

%By understanding the limits on exoplanetary
%trixaxialities, we place constraints on the possibility of resonant
%spin-orbit capture of rocky exoplanets.

%Upper limits on a planet's triaxiallity come from geophysics.  In
%order to maintain a non-hydrostatic bulge, internal stresses must
%resist the weight of the planet's triaxiality.  If internal stresses
%become too large, the rock of the planet either plastically yields or fractures.

In Section~\ref{sec:OrderOfMag} we provide an order-of-magnitude estimate of the maximum
traxial deformation $\epsilon_{\rm max}$ of a bare rocky
planet. Section~\ref{sec:QuantCalc} contains detailed calculations of $\epsilon_{\rm max}$ 
for rocky planets with and without a fluid envelope or atmosphere.  In
Section~\ref{sec:Discuss} we summarize our result and discuss its implications for
resonant spin-orbit capture and asynchronous planet rotation.

%By modeling rocky exoplanets as homogeneous, incompressible,
%self-gravitating elastic spheres, we place upper limits on the
%triaxiality of rocky exoplanets.  In Section~\ref{sec:SetupForm}, we
%describe the setup and formalism which we will use throughout the
%paper.  In Section~\ref{sec:MaxEps}, we derive the maximal ellipticity
%of a bare, rocky exoplanet.  Section~\ref{sec:EffectFluidEnv} explores
%how the maximal ellipticity changes with simple models of exoplanetary
%atmospheres and oceans.  Section~\ref{sec:Discuss} summarizes our work
%and discusses implications.

\section{Order of Magnitude Estimate of the Maximum Triaxiality of Rocky Planets}
\label{sec:OrderOfMag}

For a rocky planet of density $\rgc$ and radius $\Rc$, the anisotropic stress associated with the weight of its triaxial deformation is or order
\be
|\bT_{\rm grav}| \sim \rgc \gc \Rc \eps,
\ee
where $\gc = G M_{\rm c}/\Rc^2$ is the gravitational acceleration, and $\eps$ is the dimensionless triaxiality [defined in Eq.~\eqref{eq:eps} below]. The stress $\bT_{\rm grav}$ must be balanced by internal elastic stress.  A rough magnitude of the elastic stresses is
\be
|\bT_{\rm elast}| \sim \mu u,
\ee
where $\mu$ is the shear modulus and $u$ is the strain.  This gives
\be
u \sim \frac{\rgc \gc \Rc}{\mu} \eps.
\label{eq:sg_scale}
\ee
The planet can yield plastically or fracture when $u$ exceeds a critical value $\ucrit$ (of order $10^{-5} - 10^{-3}$).  Thus the maximum triaxiality is
\be
\epsmax \sim \left( \frac{\mu}{\rgc \gc \Rc} \right) \ucrit.
\label{eq:scale_epsmax}
\ee
Detailed calculation in Section~\ref{sec:QuantCalc} reproduces the same scaling relation except $\epsmax$ is a factor of $7.9$ larger.  Thus
\be
\epsmax \simeq 7.9 \left( \frac{\mu}{\rgc \gc \Rc} \right) \ucrit.
\ee

\section{Quantitative Calculation}
\label{sec:QuantCalc}

We model the planet to as a constant density core \footnote{  A real planet may consist of a solid/liquid core, a mantle, a crust and an liquid envelope/atmosphere. In our simple planet model, the region inside $\Rc$ (with constant density and rigidity) is termed ``rocky core"
or ``core", while the region outside $\Rc$ (with zero rigidity) termed envelope. } ($\text{density} = \rgc$) with radius $\Rc$,  with a fluid envelope extending to radius $R$.  We consider two types of envelopes:
\begin{enumerate}
\item An isothermal atmosphere, with equation of state $\rg = p/\cs^2$.
\item A constant density ocean, with $\rg = \rgo$.
\end{enumerate}
Here, $p$ is the pressure and $\cs$ is the (constant) sound speed.  We assume the atmosphere is thin, with scale height $\cs^2/\gc \ll \Rc$, where $\gc =(4\pi/3) \rgc \Rc$ is the gravitational acceleration at $r = \Rc$.  

We take the equilibrium state to be spherically symmetric with no shear stress.  The equations of hydrostatic equilibrium are
\begin{align}
-&\frac{\der p}{\der r} - \rg \frac{\der \phi}{\der r} = 0
\label{eq:hydro} \\
&\frac{1}{r^2} \frac{\der}{\der r} \left( r^2 \frac{\der \phi}{\der r} \right) = 4 \pi G \rg,
\label{eq:poisson}
\end{align}
where $\phi$ is the gravitational potential.  We require $\der \phi/\der r|_{r=0} = 0$, $p(R) = 0$, and continuity of $p$, $\phi$, and $\der \phi/\der r$ at $r=\Rc$.  For a  bare rocky planet, the solutions for $\phi$ and $p$ are (for $r<\Rc$)
\begin{align}
\phi(r) &= -\pi G \rgc \left( 2 R^2 - \frac{2}{3} r^2 \right),
\label{eq:phi_bare} \\
p(r) &= \frac{2\pi}{3} G \rgc^2 (R^2 - r^2).
\label{eq:p_bare}
\end{align}
For a planet with an isothermal atmosphere, the solutions are (to leading order in $\cs^2/\Rc \gc$)
\begin{align}
\phi(r) &= \left\{ 
\begin{array}{ll}
-\pi G \rgc (2 \Rc^2 - \frac{2}{3} r^2) & r \le \Rc \\
- \frac{4\pi}{3} G \rgc \Rc^3/r & r > \Rc
\end{array} \right. 
\label{eq:phi_atm}  ,\\
p(r) &= \left\{
\begin{array}{ll}
\frac{2\pi}{3} G \rgc^2(\Rc^2 - r^2) + \cs^2 \rga & r \le \Rc \\
\rga \cs^2  \exp[-(r-\Rc) \gc/\cs^2 ] & r > \Rc
\end{array} \right. ,
\label{eq:p_atm}
\end{align}
where $\rga$ is the density at the base of the atmosphere, and is a free parameter.  For a planet with a constant density ocean, the solutions are (for $r \le R$)
\begin{align}
\phi(r) &= \left\{ \begin{array}{ll}
-\pi G \rgc (2 \Rc^2 - \frac{2}{3} r^2) & r \le \Rc \\
- \frac{4\pi}{3} \Rc^3 (\rgc - \rgo)/r - \pi G \rgo( 2 R^2 - \frac{2}{3} r^2 ) & r > \Rc
\end{array} \right. , 
\label{eq:phi_ocean}\\
p(r) &= - \rg(r) \phi(r) + \rgo \phi(R).
\label{eq:p_ocean}
\end{align}

We then perturb the surface of the  core, so that the  core's radius is given by
\be
r_{\rm c} = \Rc + \dg \Rc \, \text{Re}[ Y_{22}(\theta, \vphi)],
\label{eq:dRc}
\ee
where $\dg \Rc \ll \Rc$, $Y_{\ell m}$ are spherical harmonics, and $\text{Re}$ denotes the real part.  The perturbed surface of the fluid envelope is
\be
r = R + \dg R \, \text{Re}[ Y_{22}(\theta, \vphi)].
\label{eq:dR}
\ee
The associated perturbation to the gravitational potential is
\be
\dg \phi = \dg \phi_{22}(r) \, \text{Re}[Y_{22}(\theta, \vphi)].
\label{eq:form_dphi}
\ee
Solving the perturbed Poisson's equation, we find that for a bare  rocky planet and a thin-atmosphere planet, $\dg \phi_{22}$ is given by (for $r\le R$)
\be
\dg \phi_{22}(r) = - \frac{4\pi}{5} \frac{G \rgc \Rc^2 \ag_{\rm c}^2 \dg \Rc}{\max(r,\Rc)},
\label{eq:dphi_bare}
\ee
and for a constant density ocean:
\be
\dg \phi_{22}(r) = \dg \phi_{\rm c}(r) + \dg \phi_{\rm o}(r),
\label{eq:dphi_ocean}
\ee
where
\begin{align}
\dg \phi_{\rm c}(r) = - &\frac{4\pi}{5} \frac{G \rgc \Rc^2 \ag_{\rm c}^2 \dg \Rc}{\max(r,\Rc)},\\
\dg \phi_{\rm o}(r) = - &\frac{4\pi}{5} G \rgo \left[ \frac{R^2 \ag^2 \dg R}{\max(r,R)} - \frac{\Rc^2 \ag_{\rm c}^2 \dg \Rc}{\max(r,\Rc)} \right]
\end{align}
are the perturbations in the gravitational potential from the perturbed  core and ocean, respectively, and
\be
\ag = \frac{\min(r,R)}{\max(r,R)},
\hspace{5mm}
\ag_{\rm c} = \frac{\min(r,\Rc)}{\max(r,\Rc)}.
\ee

Let $ I = I_{zz} \ge I_{yy} \ge I_{xx}$ be the principal components of the planet's moment of inertia tensor.  To linear order in all perturbed quantities, the triaxiality $\eps$ is
\be
\eps \equiv \frac{I_{yy} - I_{xx}}{I_{zz}} = \sqrt{ \frac{32 \pi}{15} } \frac{Q_{22}}{I},
\label{eq:eps}
\ee
where $Q_{22}$ is the second gravitational moment of the planet.  Because $Q_{22}$ is related to the perturbed gravitational potential at the surface of the planet through
\be
\dg \phi_{22}(R) = -\left( \frac{4\pi}{5} \right) \frac{G Q_{22}}{R^3},
\ee
we may write
\be
\eps = -\sqrt{ \frac{10}{3\pi} } \frac{R^3}{G I} \dg \phi_{22}(R).
\label{eq:eps_pot}
\ee
Thus, the triaxiality may be obtained by evaluating the perturbed potential at the surface of the planet.  For a planet with an isothermal atmosphere (which formally extends to infinity), we evaluate Eq.~\eqref{eq:eps_pot} at $r = \Rc$.  Corrections to Eq.~\eqref{eq:eps_pot} from the gravitational potential of the atmosphere are of order $\cs^2/(\gc \Rc) \ll 1$.

Elastic stresses are required to resist the non-isotropic weight on the  core from the planet's ellipticity.  Assuming the  core to be homogeneous ($\text{shear modulus} = \mu = \text{constant}$) and incompressible, the perturbed equations of elastostatic equilibrium in the  core are \citep{LandauLifshits(1959)}
\begin{align}
-\bdel \dg p + \mu \del^2 \bxi - \rgc \bdel \dg \phi &= 0, 
\label{eq:elasto} \\
\bdel \bcdot \bxi &= 0,
\label{eq:incomp}
\end{align}
where $\bxi$ is the Lagrangian displacement, $\dg \rg$, $\dg p$, and $\dg \phi$ are the Eulerian perturbations.  In the fluid envelope, the equations of hydrostatic equilibrium are
\be
-\bdel \dg p - \dg \rg \bdel \phi - \rg \bdel \dg \phi = 0,
\label{eq:hydro_env}
\ee
coupled with the perturbed equation of state
\be
\dg \rg = \frac{\der \rg}{\der p} \dg p.
\label{eq:EOS}
\ee
Equations~\eqref{eq:elasto}-\eqref{eq:EOS} are solved with the boundary conditions $\bxi(0) = 0$, $[-\dg p - \bxi \bcdot \bdel p]_{r=R} = 0$, and at the  core-envelope boundary ($r=\Rc$), we require continuity of $\vr \bcdot \bxi$ and the Lagrangian perturbed radial traction
\be
\Del {\bm T} = -\dg p \vr + \mu \vr \bcdot \left[ (\bdel \bxi) + (\bdel \bxi)^T \right] - (\bxi \bcdot \bdel p) \vr.
\label{eq:DT}
\ee

The strain tensor for an incompressible material is
\be
\bu = \frac{1}{2} \left[(\bdel \bxi) + (\bdel \bxi)^T \right].
\ee
We define the the strain amplitude $u$ via
\be
u^2 \equiv \frac{1}{2} \text{Tr}(\bu \bcdot \bu) ,
\label{eq:sg}
\ee
where $\text{Tr}({\bf U})$ denotes the trace of the tensor ${\bf U}$.  When $u$ exceeds a critical value $\ucrit$, the rocky  planet no longer behaves elastically, and begins to either plastically deform or fracture (the von Mises yield criterion, see \citealt{TurcotteSchubert(2002)}).  The critical strain $\ucrit$ is a material property of the  rocky planet (more specifically, the planet's mantle), and is related to the yield stress $Y$ via $\ucrit = Y/\mu$.  Laboratory studies of the strength of rocks which make up the Earth's crust \citep{Kohlstedt(1995)} and theoretical arguments on the initiation of subduction by plastic yielding in the earth's lithosphere \citep{Fowler(1993),TrompertHansen(1998),WongSolomatov(2015)} give estimates of $Y = 10^7-10^9 \, \text{dyn}/\text{cm}^2$ for the earth's lithosphere.  The characteristic shear modulus value for the  Earth is $\mu = 10^{12} \, \text{dyn}/\text{cm}^2$ \citep{TurcotteSchubert(2002)}, thus the critical strain is in the range $u_{\rm crit} = 10^{-5} - 10^{-3}$.

The strain required to resist the anisotropic weight of a triaxial  planet is non-uniform, and assumes a maximum (peak) value $\upeak$ at a certain location  in the planet.  When $\upeak$ exceeds $\ucrit$, the  core either plastically deforms or fractures, reducing the strain in the  surface and core, and hence reducing $\eps$.  Therefore, the maximal triaxiality $\eps_{\rm max}$ of the rocky  planet is set by $\upeak = \ucrit$.

\subsection{Bare Rocky Planet}
\label{sec:BareCore}

We show in the Appendix that the solution $\bxi$ of Eqs.~\eqref{eq:elasto}-\eqref{eq:incomp} may be written in the form
\be
\bxi = \xir(r) \vr \, \text{Re}[Y_{22}(\theta,\vphi)] + \xip(r) \, \text{Re}[ r \bdel Y_{22}(\theta,\vphi)],
\label{eq:bxi}
\ee
where
\begin{align}
\xir(r) &= \xi_1 \left( \frac{r}{\Rc} \right)^3 + 2 \xi_3 \left( \frac{r}{\Rc} \right),
\label{eq:xir} \\
\xip(r) &= \xi_2 \left( \frac{r}{\Rc} \right)^3 + \xi_3 \left( \frac{r}{\Rc} \right),
\label{eq:xip}
\end{align}
and $\xi_1, \xi_2$, and $\xi_3$ are constants.  Applying the boundary conditions at $r = 0$ and $r = \Rc$, we find
\begin{align}
\xi_1 &= \frac{6}{95} \left( \frac{\rgc \gc \Rc}{\mu} \right) \dg \Rc,
\label{eq:xi1_BareCore} \\
\xi_2 &= \frac{1}{19} \left( \frac{\rgc \gc \Rc}{\mu} \right) \dg \Rc,
\label{eq:xi2_BareCore} \\
\xi_3 &= - \frac{8}{95} \left( \frac{\rgc \gc \Rc}{\mu} \right) \dg \Rc
\label{eq:xi3_BareCore}.
\end{align}
From Eq.~\eqref{eq:sg}, the corresponding strain amplitude $u$ is given by
\be
u^2= \frac{1}{2}u_{rr}^2 + \frac{1}{2}u_{\theta\theta}^2 + \frac{1}{2}u_{\vphi\vphi}^2 + u_{\theta\vphi}^2 + u_{r\theta}^2 + u_{\vphi r}^2,
\label{eq:barsg2_form}
\ee
where \citep{LandauLifshits(1959)}
\begin{align}
u_{rr} &= A_{22}\frac{\der \xir}{\der r} \sin^2 \theta \cos2\vphi, \\
u_{\theta \theta} &= A_{22} \left[ \frac{\xir}{r} \sin^2\theta + \frac{2 \xip}{r} (\cos^2 \theta - \sin^2\theta) \right]\cos 2\vphi, \\
u_{\vphi\vphi} &= A_{22} \left[ \frac{\xir}{r} \sin^2 \theta - \frac{2\xip}{r} (\sin^2\theta+1) \right]\cos2\vphi, \\
u_{\theta\vphi} &= -6A_{22} \frac{\xip}{r} \cos\theta \sin2\vphi, \\
u_{r\theta} &= A_{22} \left( \frac{\xir}{r} + \frac{\der \xip}{\der r} - \frac{\xip}{r} \right) \sin\theta \cos\theta \cos2\vphi, \\
u_{\vphi r} &= -A_{22} \left( \frac{\xir}{r} + \frac{\der \xip}{\der r} - \frac{\xip}{r} \right) \sin\theta \sin2\vphi,
\end{align}
and $A_{22} = \sqrt{15/32\pi}$.  As expected, the strain amplitude $u$ scales as $\rgc \gc \dg \Rc/\mu$.  Therefore, we define the rescaled strain magnitude
\be
\baru \equiv \left( \frac{\mu}{\rgc \gc \dg \Rc} \right) u.
\label{eq:barsg}
\ee

\begin{figure}
\centering
\includegraphics[scale=0.45]{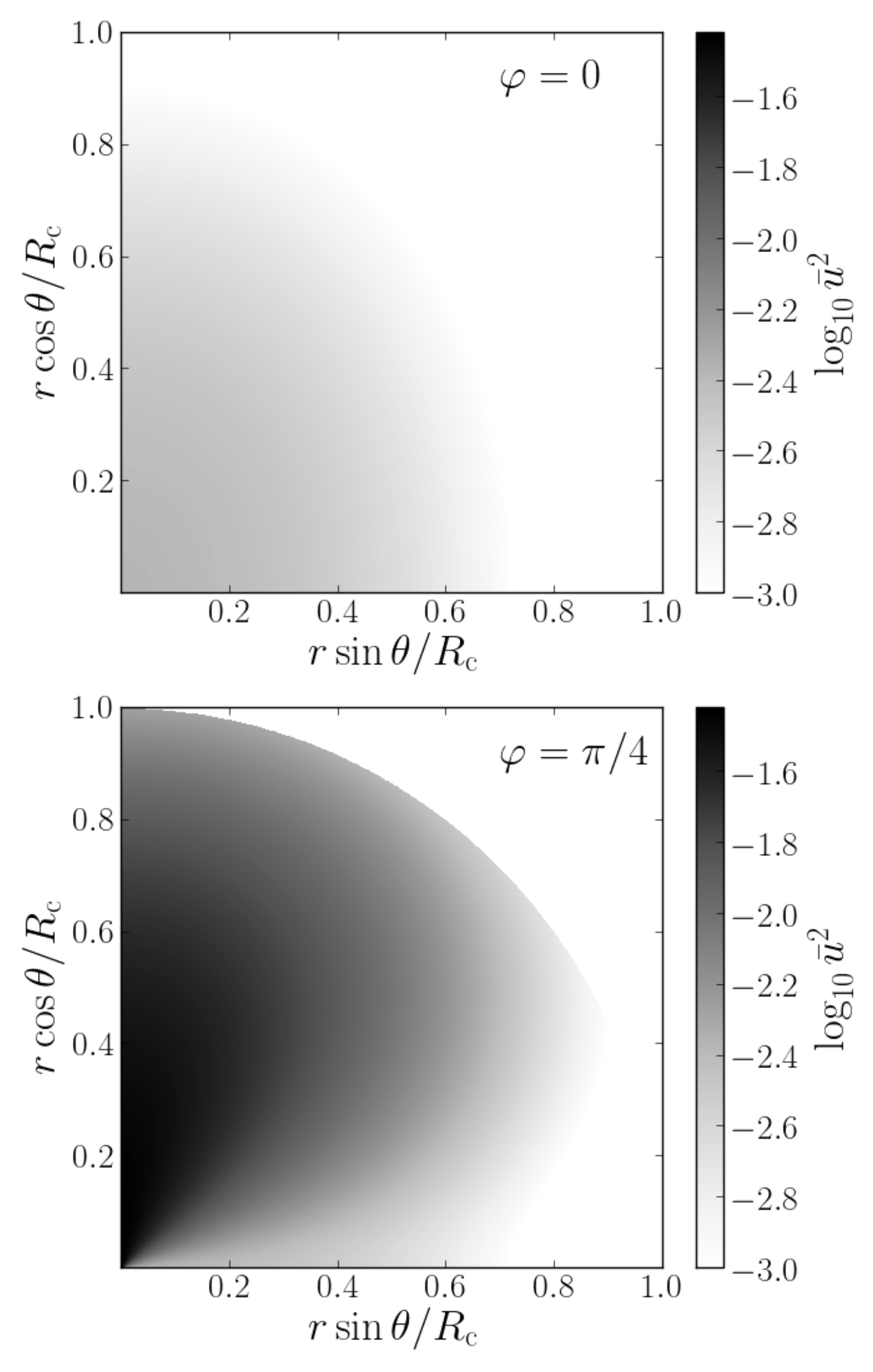}
\caption{
Rescaled strain magnitude $\baru^2$ [Eq.~\eqref{eq:barsg}] for a bare rocky planet as a function of the coordinates $(r,\theta)$.  The values of azimuth $\vphi$ are as indicated.}
\label{fig:barsg2_BareCore}
\end{figure}

In Figure~\ref{fig:barsg2_BareCore}, we plot the rescaled $\baru^2$ over coordinates $(r,\theta)$, for $\vphi = 0$ and $\vphi = \pi/4$.   Deep in the planetary interior ($r \sim 0.0-0.2 \Rc$) is where the planetary strain is highest, with a maximal value of
\be
\barupeak \equiv \max_{r \in [0,\Rc]}\big[\baru(r,\theta,\vphi)\big] = 0.195.
\ee
The source of this strain is not the direct response to the weight of the planet's triaxiality, which scales with radius as $\baru^2 \propto r^4$ [$\xi_1$ and $\xi_2$ terms in Eqs.~\eqref{eq:xir}-\eqref{eq:xip}].   Instead, the strain deep in the planetary interior comes from additional stresses to make the radial traction [Eq.~\eqref{eq:DT}] vanish on the planet's surface, which scales with radius as $\baru^2 \propto \text{constant}$ [$\xi_3$ terms in Eqs.~\eqref{eq:xir}-\eqref{eq:xip}].  In comparison, we find the maximal rescaled strain on the planetary surface $\barusurf$ to be
\be
\barusurf \equiv \max_{r=\Rc}\big[\bar u(r,\theta,\vphi)] = 7.32 \times 10^{-2}.
\ee
 The value $\barupeak$ differs from $\barusurf$ by a factor of $\sim 3$.  There is some uncertainty as to what is the correct location one should equate $u$ with $\ucrit$ to obtain the planet's maximal triaxiality.  For instance, a substantial portion of the Earth's core is fluid \citep{TurcotteSchubert(2002)}, so it is unable to sustain any anisotropic strain $u$.  Due to the crudeness of our model, we still equate $\ucrit$ with $\upeak$ to calculate $\epsmax$, and note that realistic equations of state  for terrestrial planets may change this result by factors of order unity.

From Eqs.~\eqref{eq:dphi_bare} and~\eqref{eq:eps_pot}, we have $\eps = \sqrt{15/2\pi} (\dg \Rc/\Rc)$, thus
\be
\upeak \simeq \frac{1}{7.9} \frac{\rgc \gc \Rc}{\mu} \eps.
\label{eq:sgpeak}
\ee
Equating $\upeak$ with $\ucrit$ gives the maximum triaxiality of the bare rocky planet:
\begin{align}
\epsmax \simeq \ &7.9 \frac{\mu}{\rgc \gc \Rc} \ucrit
\nonumber \\
\simeq \ &1.9 \times 10^{-5}  \left( \frac{\mu}{10^{12} \, \text{dyn}/\text{cm}^2} \right) 
\nonumber  \\
&\times\left( \frac{\rg}{6 \, \text{g}/\text{cm}^3} \right)^{-2} \left( \frac{R}{R_\oplus} \right)^{-2} \left( \frac{\ucrit}{10^{-5}} \right).
\label{eq:epsmax_BareCore}
\end{align}

\subsection{Planet with a Thin Isothermal Atmosphere}
\label{sec:ThinIso}

Applying the boundary conditions at the  core-envelope boundary, we find
\begin{align}
\xi_1 &= \frac{6}{95} \left( 1 - \frac{\rga}{\rgc} \right) \left( \frac{\rgc \gc  \Rc}{\mu} \right)\dg \Rc,
\label{eq:xi1_Isotherm} \\
\xi_2 &= \frac{1}{19} \left( 1 - \frac{\rga}{\rgc} \right) \left( \frac{\rgc \gc \Rc}{\mu} \right)\dg \Rc,
\label{eq:xi2_Isotherm} \\
\xi_3 &= - \frac{8}{95} \left( 1 - \frac{\rga}{\rgc} \right) \left( \frac{\rgc \gc \Rc}{\mu} \right)\dg \Rc
\label{eq:xi3_Isotherm}.
\end{align}
where $\rga$ is the density at the base of the atmosphere [see Eq.~\eqref{eq:p_atm}].  Clearly, unless $\rga \sim \rgc$, the reduced strain in the  core from the addition of a thin isothermal atmosphere is negligible.  The peak strain amplitude in the  core is
\be
\upeak \simeq \frac{1}{7.9} \frac{\rgc \gc \Rc}{\mu} \eps \left( 1 - \frac{\rga}{\rgc} \right).
\label{eq:barsgmax_Isotherm}
\ee
The maximum triaxiality is larger than Eq.~\eqref{eq:epsmax_BareCore} by the factor $(1 - \rga/\rgc)^{-1}$.

\subsection{Planet with a Constant Density Ocean}
\label{sec:ConstDenOcean}

\begin{figure}
\centering
\includegraphics[scale=0.5]{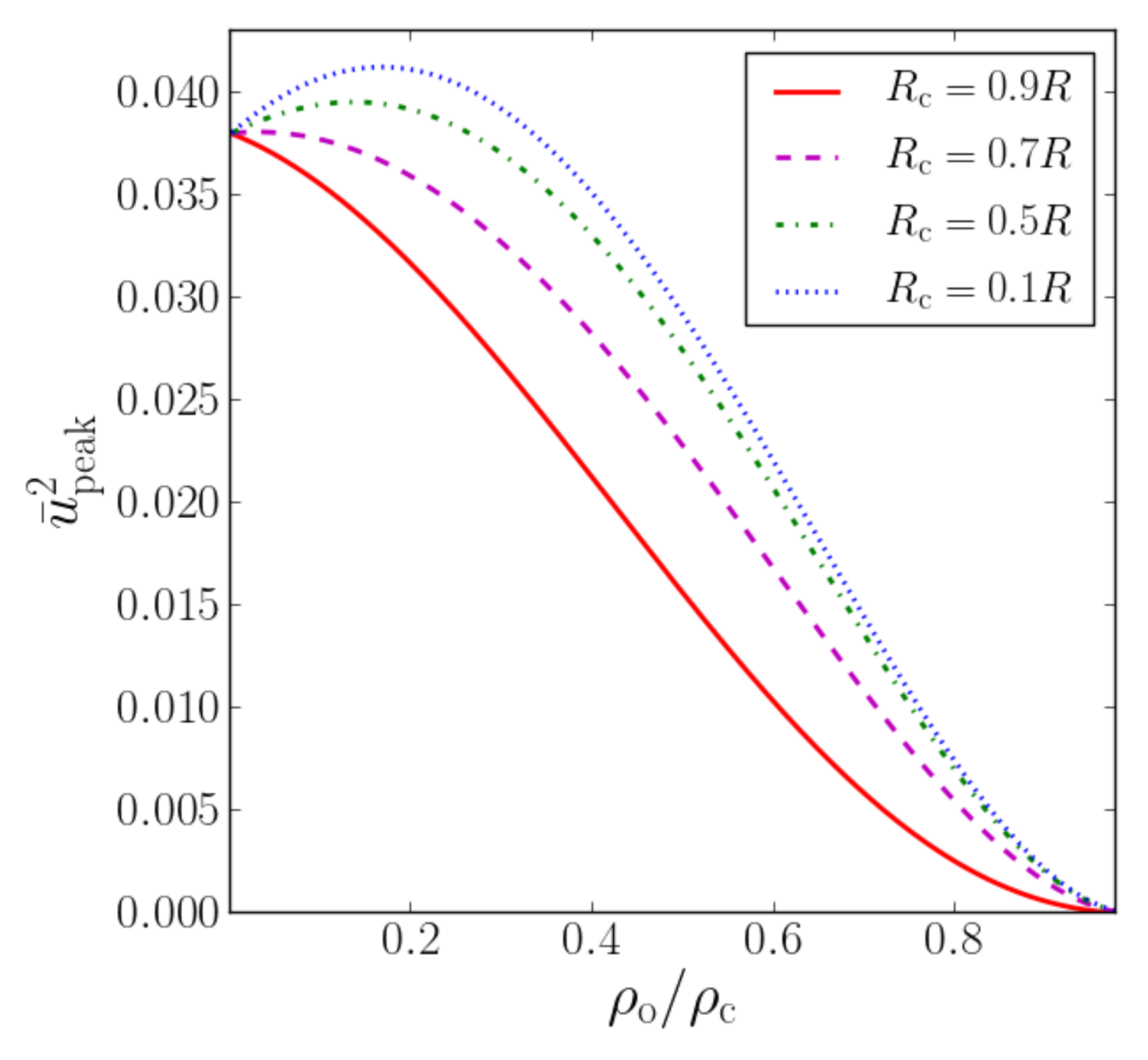}
\caption{
The square of the surface strain amplitude $\barupeak$ for a rocky  core under a constant density ocean [Eq.~\eqref{eq:barsgmax_ConstDenOcean}], plotted as a function of $\rgo/\rgc$, with values of $\Rc/R$ as indicated.  Here, $\rgo = \text{ocean density}$, $\rgc = \text{ core density}$, $\Rc = \text{ core radius}$, and $R = \text{ envelope radius}$.  }
\label{fig:barsg2_ConstDenOcean}
\end{figure}

When the  core is surrounded by a constant density ocean, we apply the boundary conditions at the  core-envelope interface and find
\begin{align}
\xi_1 &= \frac{6}{95} F_1 \left( \frac{\Rc}{R} , \frac{\rgo}{\rgc} \right) \left( \frac{\rgc \gc \Rc}{\mu} \right) \dg \Rc,
\label{eq:xi1_ConstDenOcean} \\
\xi_2 &= \frac{1}{19} F_1 \left( \frac{\Rc}{R} , \frac{\rgo}{\rgc} \right) \left( \frac{\rgc \gc \Rc}{\mu} \right)\dg \Rc,
\label{eq:xi2_ConstDenOcean} \\
\xi_3 &= - \frac{8}{95} F_1 \left( \frac{\Rc}{R} , \frac{\rgo}{\rgc} \right) \left( \frac{\rgc \gc \Rc}{\mu} \right) \dg \Rc
\label{eq:xi3_ConstDenOcean},
\end{align}
where
\be
F_1 (x,y) = \left( 1 + \frac{3 y}{2} \right)(1-y) - \frac{9 y x^5 (1-y)}{10 [ x^3 (1-y) + 2y/5]}.
\ee
Notice that $F_1(\Rc/R,0) = 1$, reproducing the results of Section~\ref{sec:BareCore}, and also $F_1(\Rc/R,1) = 0$, showing that when $\rgo = \rgc$, the ocean's weight completely cancels the weight from the  planet's ellipticity.  Eqs.~\eqref{eq:xi1_ConstDenOcean}-\eqref{eq:xi3_ConstDenOcean} give the rescaled peak strain amplitude of 
\be
\barupeak = \frac{\mu}{\rgc \gc \dg \Rc} \upeak = 0.195 F\left( \frac{\Rc}{R} , \frac{\rgo}{\rgc} \right).
\label{eq:barsgmax_ConstDenOcean}
\ee

Plotted in Figure~\ref{fig:barsg2_ConstDenOcean} is the square of Eq.~\eqref{eq:barsgmax_ConstDenOcean} as a function of $\rgo/\rgc$, with values of $\Rc/R$ as indicated.  We see that when the  core is small ($\Rc \lesssim 0.5 R$) with a low density ocean ($\rgo \lesssim 0.2 \rgc$), the presence of an ocean  increases the strain in the  core.  When the  core is large ($\Rc \gtrsim 0.5 R$) or the ocean is dense ($\rgo \gtrsim 0.2 \rgc$), the weight of the ocean works to cancel the weight on the  core from the  planet's triaxiality $\eps$.

\begin{figure}
\centering
\includegraphics[scale=0.5]{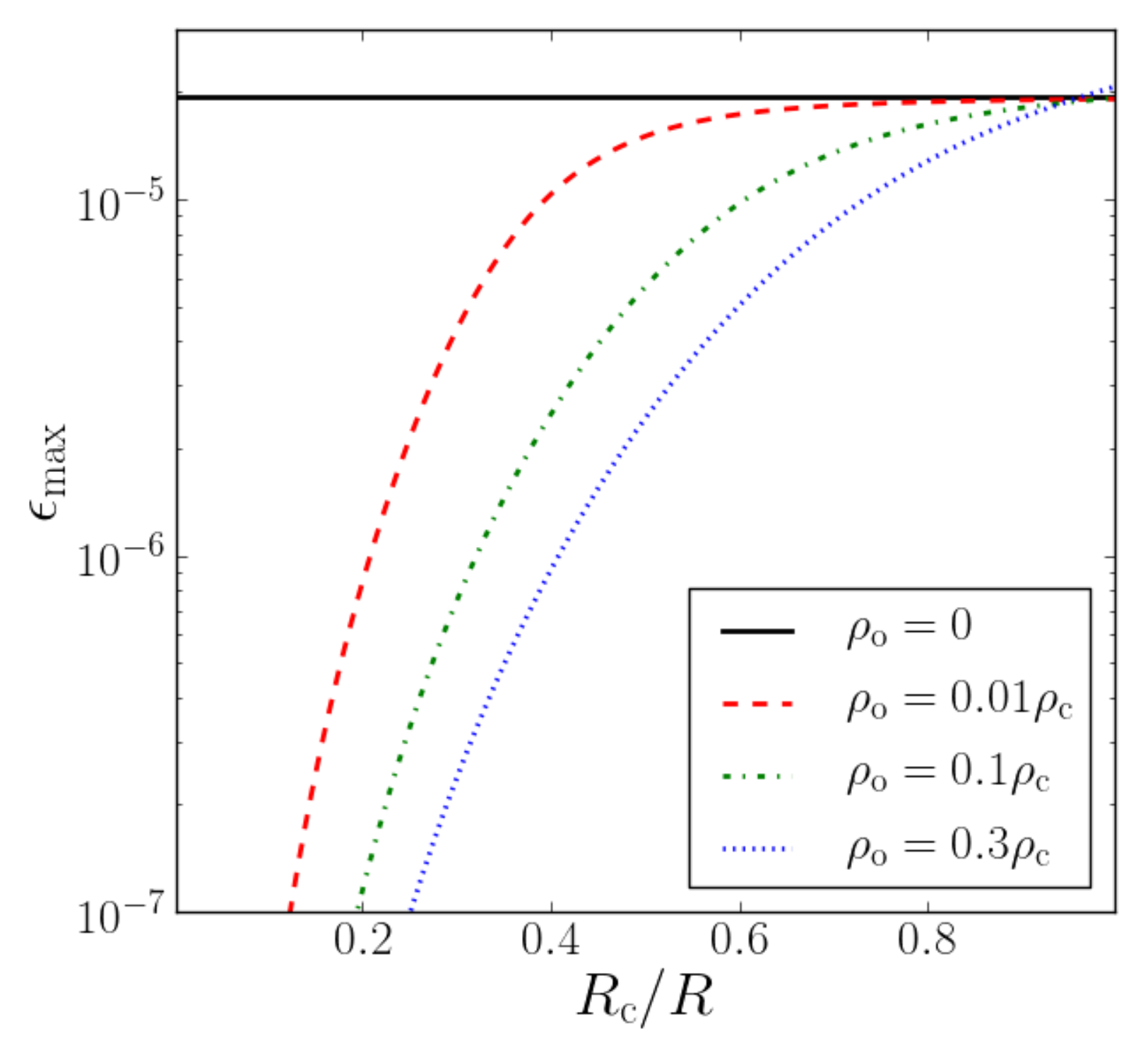}
\caption{
Maximal Triaxiallity $\epsmax$ for a planet with a constant density ocean as a function of $\Rc/R$, with values of $\rgo/\rgc$ as indicated.  Here, $\rgo = \text{ocean density}$, $\rgc = \text{ core density}$, $\Rc = \text{ core radius}$, and $R = \text{ envelope radius}$.  We take $\mu = 10^{12} \, \text{dynes}/\text{cm}^2$, $\rgc = 6 \, \text{g}/\text{cm}^3$, $\Rc = R_\oplus$, and $\ucrit = 10^{-5}$.}
\label{fig:epsmax_ConstDenOcean}
\end{figure}

With a constant density ocean, $\eps$ and $\dg \Rc$ are related by
\be
\eps = \sqrt{ \frac{15}{2\pi} } F_2 \left( \frac{\Rc}{R} , \frac{\rgo}{\rgc} \right) \frac{\dg \Rc}{\Rc},
\label{eq:eps_dRc_ConstDenOcean}
\ee
where
\be
F_2(x,y) = \frac{1-y}{1 - y + y/x^5} \left[ 1 + \frac{3 y}{5x^3(1-y) + 2y} \right].
\ee
Notice that $F_2(\Rc/R,0) = 1$, recovering the bare  rocky planet result.  Also note that $F_2(\Rc/R,1) = 0$, showing when $\rgo = \rgc$ the planet has no triaxiality.  Using Eqs.~\eqref{eq:barsgmax_ConstDenOcean} and~\eqref{eq:eps_dRc_ConstDenOcean}  and setting $\upeak = \ucrit$, we obtain the maximum triaxiality
\begin{align}
\epsmax = 7.9 \frac{\mu}{\rgc \gc \Rc} \ucrit F \left( \frac{\Rc}{R}, \frac{\rgo}{\rgc} \right),
\label{eq:epsmax_ConstDenOcean}
\end{align}
where
\be
F(x,y) = F_2(x,y)/F_1(x,y).
\ee

Figure~\ref{fig:epsmax_ConstDenOcean} shows $\epsmax$ [Eq.~\eqref{eq:epsmax_ConstDenOcean}] as a function of $\Rc/R$, with values of $\rgo/\rgc$ as indicated.  We see that the maximal triaxiality of the planet may be significantly decreased by the presence of an ocean.  This is because as $\Rc/R \to 0$, the bulge at the planetary surface induced by the  planet's triaxiality becomes increasingly negligible, even though the strain in the  core may be reduced by the presence of an ocean (see Fig.~\ref{fig:barsg2_ConstDenOcean}).

\section{Summary and Discussion}
\label{sec:Discuss}

%\subsection{Summary}                                                                        
%\label{sec:Summ}                                                                            

We have derived the analytic expression for the maximum triaxial
deformation $\epsmax$ of a rocky planet as a function of its
density and radius [see Eq.~\eqref{eq:epsmax_BareCore}]. This maximum triaxiality depends on
the rigidity (shear modulus) and critical strain of the rocky
material. A thin atmosphere surrounding the rocky  core has a negligible
impact on $\epsilon_{\rm max}$ [see Eq.~\eqref{eq:barsgmax_Isotherm}], while a liquid ocean
envelope may lower the maximal triaxiality by a factor of a few or more
than an order of magnitude, depending on the thickness and density of
the ocean [see Eq.~\eqref{eq:epsmax_ConstDenOcean} and Figs.~\ref{fig:barsg2_ConstDenOcean}-\ref{fig:epsmax_ConstDenOcean}].

%In this work, by modelling exoplanets as homogeneous, incompressible, self-gravitating elastic spheres, we have put a simple analytic limit to the maximal triaxiality of a rocky exoplanet (Sec.~\ref{sec:MaxEps}).  The scaling may be understood from an order of magnitude argument (Sec.~\ref{sec:OrderOfMag}), while a detailed calculation gives the maximal triaxiality~\eqref{eq:epsmax}.  A thin isothermal atmosphere has a negligible impact on the maximal triaxiality (Sec.~\ref{sec:ThinIso}), while an ocean mainly works to lower the maximal elipticity (Sec.~\ref{sec:ConstDenOcean}).  Therefore, the rough magnitude given by Eq.~\eqref{eq:epsmax} is unlikely to change with more detailed physical modelling of rocky exoplanets.  

%\subsection{Implications}
%\label{sec:Imps}

\begin{table}
\centering
\begin{tabular}{ |c|c|c| }
\hline
Body & Observed $\eps$ & $\epsmax$ \\
\hline
Mercury & $1.3 \times 10^{-4}$ & $1.6 \times 10^{-4}$ \\
Venus & $5.4 \times 10^{-6}$ & $2.9 \times 10^{-5}$ \\
Earth & $1.5 \times 10^{-5}$ & $2.3 \times 10^{-5}$ \\
Mars & $5.5\times 10^{-4}$ & $1.6 \times 10^{-4}$ \\
Moon & $2.3\times 10^{-4}$ & $8.7 \times 10^{-4}$ \\
\hline
\end{tabular}
\caption{The observed traxiality $\eps$ 
compared to $\epsmax$ computed with Eq.~\protect\eqref{eq:epsmax_BareCore} for several
terrestrial bodies in the solar system. The observed $\eps$ is related to 
the gravity field coefficient by $\epsilon=10 C_{22}$ (we assume $I = 2 M_{\rm p} R^2/5$), and is listed 
in \protect\cite{Yoder(1995)}, while Mercury's $C_{22}$ is from \protect\cite{Smith(2012)}.  
We take $\mu = 10^{12} \, {\rm dyn}/{\rm cm}^2$ and $u_{\rm crit} = 10^{-5}$
for all bodies.}
\label{tab:triaxiality}
\end{table}

The rigidity $\mu$ and critical strain $u_{\rm crit}$ for rocky
planets are unknown. The value of $u_{\rm crit}$ is particularly
uncertain, and probably depends on the assembly history of the planet.
Applying our result to terrestrial bodies in the solar system, we find
that the observed values of $\epsilon$ are consistent with 
our predicted $\epsilon_{\rm max}$ (see Table~\ref{tab:triaxiality}) for reasonable $\mu$
($\sim 10^{12}$~dynes/cm$^2$) and $u_{\rm crit}$
($10^{-5}-10^{-3}$). Interestingly, for Mercury, Earth, Mars,  and the Moon, these observed $\epsilon$ values are 
close to $\epsilon_{\rm max}$ with $u_{\rm crit}=10^{-5}$, the lower 
range of the critical strain for the Earth.

%In Table~\ref{tab:triaxiality}, we compare observed values of the triaxiality $\eps$ [Eq.~\eqref{eq:eps}] to the maximal value of the triaxiality for a rocky self-gravitating body $\epsmax$ [Eq.~\eqref{eq:epsmax}].  Observed values of $\eps$ are computed through \citep{Yoder(1995)}
%\be \eps \approx \frac{I_{yy} - I_{xx}}{2 M R^2/5} = 10 \, C_{22}.
%\label{eq:epsobs} \ee
%Here, $C_{ij}$ are the gravity field coefficients and $M$ is the mass of the body.  Table~\ref{tab:triaxiality} shows that in the solar system, the observed value of $\eps$ is typically 2-3 orders of magnitude lower than $\epsmax$.  We do not compare observed values of $\eps$ to $\epsmax$ for the irregular satellites (such as Hyperion), as $\epsmax$ for the irregular satellites exceeds unity, and our perturbative treatment breaks down.

\subsection{Implication for Spin-Orbit Resonance Capture}
\label{sec:Imps}

As noted in Section 1, tidal dissipation tends to drive a close-in planet toward synchronous
rotation.  The magnitude of the tidal torque on the planet reads
\be
T_{\rm tide} \simeq \frac{3 G M_\star^2 R^5}{2a^6} \frac{k_2}{Q},
\ee
where $k_2$ and $Q$ are the Love number and tidal quality factor
of the planet, respectively. This gives the tidal synchronization time
\begin{align}
t_{\rm sync}={I\Omega\over T_{\rm tide}}&=
3.5\times 10^5\,\left(\frac{Q/k_2}{ 10^3}\right)\left(\frac{\rgc}{ 6\,{\rm g/cm}^3}\right)
\nonumber \\
&\times \left(\frac{M_\star}{ 0.3 \, M_\odot}\right)^{-3/2}\left(\frac{a}{ 0.1\,\text{AU}}\right)^{9/2}\,
\text{years},
\label{eq:tsync}
\end{align}
where $M_\star$ is the host star mass, $a$ is the planetary semi-major axis, and $\Om \simeq \sqrt{GM_\star /a^3}$ is the planetary orbital angular frequency.
We have scaled $a$ to 0.1~AU, the characteristic HZ distance for $0.3 M_\odot$ M-dwarfs (e.g. \citealt{Shields(2016)}).
On the other hand, the tidal circularization time of the orbit is
\begin{align}
t_{\rm circ}= \ & \frac{M_{\rm p} a^2\Omega}{T_{\rm tide}} 
\nonumber \\
= \, &4.8\times 10^{12} \left( \frac{M_\star}{0.3 \, M_\odot} \right)^{-3/2} \left( \frac{a}{0.1 \, \text{AU} } \right)^{13/2}
\nonumber  \\
&\times \left( \frac{\rgc}{6 \, \text{g}/\text{cm}^3} \right) \left( \frac{R}{R_\oplus} \right)^{-2} \left( \frac{Q/k_2}{10^3} \right) \text{years},
\label{eq:tcirc}
\end{align}
where $M_{\rm p}$ is the mass of the planet.  The planet can retain its initial (``primordial'') eccentricity at formation
if $t_{\rm circ}$ is longer than the age of the system.
With a finite eccentricity, the planet may be captured into the 3:2
spin-orbit resonance during its tidal spin-down \citep{GoldreichPeale(1966)}. 
The resonance torque on the planet due to its intrinsic triaxial deformation
has a magnitude (for the 3:2 resonance)
\be
T_{\rm tri} \simeq \frac{21}{4} \eps I e \left( \frac{G M_\star}{a^3} \right).
\ee
 If the planet's rheology is a frequency-independent constant-$Q$ tidal model, a necessary condition for resonance capture is $T_{\rm tri}>T_{\rm tide}$, giving 
\begin{align}
\eps > 1.0 \times 10^{-6} &\left( \frac{M_\star}{0.3 \, M_\odot} \right) \left( \frac{a}{0.1 \, \text{AU}} \right)^{-3} 
\nonumber \\
&\times \left( \frac{\rg}{6 \, \text{g}/\text{cm}^3} \right)^{-1} \left( \frac{k_2/Q}{10^{-3}} \right) \left( \frac{e}{0.01} \right)^{-1},
\label{eq:epsmin}
\end{align}
where we have scaled the planetary eccentricity $e$ to 0.01, characteristic of super-Earth systems
discovered by the {\it Kepler} mission \citep{WuLithwick(2013),HaddenLithwick(2016)}.   Although we have assumed a simple constant-$Q$ model for the planet's rheology, one may obtain a similar lower bound for $\eps$ using an Andrade model, as long as the planet's eccentricity is low enough \citep{Ribas(2016)}.  We also note that planets with eccentricities of order $0.01$ have low probabilities for capture into 3:2 spin-orbit resonances, regardless of the rheology \citep{MurrayDemott(2000),MakarovB(2012)}. 
%DL: add REFs:
% Wu, Y., \& Lithwick, Y. 2013, ApJ, 772, 74
% Hadden, S., \& Lithwick, Y. 2016, arXiv:1611.03516

To avoid chaotic spin behavior associated with the overlap
of the synchronous and $3:2$ resonances, the planet's triaxiality must satisfy 
\citep{Wisdom(1984)}
\be
\eps < \frac{1}{3} \left( \frac{1}{2 + \sqrt{14 e}} \right)^2 \simeq \frac{1}{12}.
\label{eq:epsch}
\ee
Comparing (\ref{eq:epsmin}) and (\ref{eq:epsch}) to our derived $\eps_{\rm max}$,
we see that capture into stable asynchronous spin-orbit resonance is a distinct possibility
for planets in the HZ of M-dwarfs.

If we take $M_\star=0.08M_\odot$ and $a=0.03$~AU, appropriate for the ``habitable" planets around TRAPPIST-1 \citep{Gillon(2017)}, then the numerical factor in front of  Eq.~\eqref{eq:tcirc} becomes $1.4\times 10^{10}$ years, and that of Eq.~\eqref{eq:epsmin} becomes $10^{-5}$. This suggests that tidal dissipation cannot damp the planet's eccentricity, and the planet can be sufficiently triaxial to allow for capture into the 3:2 resonance.

\section*{Acknowledgments}

DL thanks Kevin Heng for asking the question about asynchronous planet rotation.  This work has been supported in part by NASA grants NNX14AG94G and
NNX14AP31G, and a Simons Fellowship from the Simons Foundation.  JZ is
supported by a NASA Earth and Space Sciences Fellowship in
Astrophysics.

\section*{Appendix}

To solve equations~\eqref{eq:elasto}-\eqref{eq:incomp} for an incompressible planet, we note that $\dg p$ satisfies $\del^2 \dg p = 0$.  This, with the form of $\dg \phi$ [Eq.~\eqref{eq:form_dphi}] and the requirement that $\dg p$ be finite at $r = 0$, implies that $\dg p = \dg p_{22}(r) \text{Re}[Y_{22}(\theta,\vphi)]$, with $\dg p_{22} \propto r^2$.

Define
\be
\dg v \equiv \dg p + \rgc \dg \phi = V \left( \frac{r}{\Rc} \right)^2 \text{Re}[Y_{22}(\theta,\vphi)],
\ee
where $V$ is an undetermined constant.  We decompose $\bxi$ into radial and tangential components:
\be
\bxi = \xir(r) \vr \, \text{Re}[Y_{22}(\theta,\vphi)] + \xip(r) \text{Re} [r \bdel Y_{22}(\theta,\vphi)].
\ee
Equations~\eqref{eq:elasto}-\eqref{eq:incomp} then become
\begin{align}
0 = &- \frac{2 V}{\Rc} \left( \frac{r}{\Rc} \right) + \mu \left[ \frac{\der^2 \xir}{\der r^2} + \frac{2}{r} \frac{\der \xir}{\der r} - 8 \frac{\xir}{r^2} + 12 \frac{\xip}{r^2} \right] 
\label{eq:inhomo_1} \\
0 = &- \frac{V}{\Rc} \left( \frac{r}{\Rc} \right) + \mu \left[ \frac{2 \xir}{r^2} + \frac{\der^2 \xip}{\der r^2} + \frac{2}{r} \frac{\der \xip}{\der r} - 6 \frac{\xip}{r^2} \right],
\label{eq:inhomo_2} \\
0 = & \frac{\der \xir}{\der r} + 2 \frac{\xir}{r} - 6 \frac{\xip}{r}.
\label{eq:incomp_proj}
\end{align}
Solutions to the inhomogeneous Eqs.~\eqref{eq:inhomo_1}-\eqref{eq:inhomo_2} require $\xir, \xip \propto r^3$.  Specifically, taking
\be
\xir = \xi_1 \left( \frac{r}{\Rc} \right)^3,
\hspace{5mm}
\xip = \xi_2 \left( \frac{r}{\Rc} \right)^3,
\label{eq:xi_1}
\ee
Eqs.~\eqref{eq:inhomo_1}-\eqref{eq:incomp_proj} are satisfied if
\begin{align}
0 &= - V + 2 \mu \left( \frac{\xi_1}{\Rc} + 3 \frac{\xi_2}{\Rc} \right), \\
0 &= 5 \xi_1 - 6 \xi_2.
\end{align}
In addition, we may add to Eq.~\eqref{eq:xi_1} any solutions to the equations
\be
\bdel \bcdot \bxi = 0,
\hspace{5mm}
\del^2 \bxi = 0.
\ee
It is sufficient to take
\be
\bxi = \Rc \bdel \left[ \xi_3 \left( \frac{r}{\Rc} \right)^2 \text{Re}(Y_{22}) \right].
\label{eq:xi_3}
\ee
Thus the general solutions to Eqs.~\eqref{eq:inhomo_1}-\eqref{eq:incomp_proj} take the forms
 \begin{align}
 \xir(r) &= \xi_1 \left( \frac{r}{\Rc} \right)^3 + 2 \xi_3 \left( \frac{r}{\Rc} \right), \\
 \xip(r) &= \xi_2 \left( \frac{r}{\Rc} \right)^3 + \xi_3 \left( \frac{r}{\Rc} \right).
 \end{align}
  These equations are completed by requiring continuity of $\xir$ and the radial traction $\Delta {\bm T}$ [Eq.~\eqref{eq:DT}] at the planet-envelope boundary ($r = \Rc$).  Specifically, we require 
 \begin{align}
  &\xir(\Rc^-) = \xir(\Rc^+), \\
  &\left[ - \dg p_{22} - \xir \frac{\der p}{\der r} + 2\mu \frac{\der \xir}{\der r} \right]_{r=\Rc^-}
   = \left[ - \dg p_{22} - \xir \frac{\der p}{\der r} \right]_{r=\Rc^+}, \\
  &\left[ \frac{\xir}{r} + \frac{\der \xip}{\der r} - \frac{\xip}{r} \right]_{r=\Rc^-} = 0.
\end{align}

\end{document}